THE DISCOVERY, DISCLOSURE, AND INVESTIGATION OF CVE-2024-25825

BY

HUNTER CHASENS

A Thesis

Submitted to the Division of Natural Science
New College of Florida
in fulfillment of the requirements for the degree Computer Science
Bachelor of Arts
Under the sponsorship of David Gillman

Sarasota, Florida
May, 2024

# Acknowledgments

First and foremost I'd like to thank my wonderful cat Mazi. She has not supported me in any way nor has she done anything but annoy me while I worked on this. Her greatest contributions weren't in the typos she added while walking on my keyboard, nor were they the many distractions she dragged onto my desk, but the love she showed me every morning by purring loudly on my chest as I would wake. I'd also like to thank all my computer science professors who've helped me get to this point, David Gillman, Matthew Lepinski, Tania Roy, Caitrin Eaton, Fahmida Hamid, along with all my other, non-computer science, professors who helped me get this far. Finally I'd like to thank my friends who've been there for me and my family who've supported me throughout this process

THE DISCOVERY, DISCLOSURE, AND INVESTIGATION OF CVE-2024-25825

Hunter Chasens

New College of Florida, 2024

**Abstract**

CVE-2024-25825 is a vulnerability found in FydeOS. This thesis describes its discovery, disclosure, and its further investigation in connection to a nation state actor. The vulnerability is CWE-1392: Use of Default Credentials, CWE-1393: Use of Default Password, and CWE-258: Empty Password in Configuration File found in the /etc/shadow configuration file. The root user's entry in the /etc/shadow file contains a wildcard allowing entry with any, or no, password. Following responsable disclosure, Fyde, CISA, and Mitre were informed. Fyde was already aware of the vulnerability. There was concern that this vulnerability might have been purposefully placed, perhaps by a nation state actor. After further investigation, it apears that this is unlikely to be the case. In cases in which poisoned code is suspected it might be prudent to contact the appropriate CERT, rather than the parent company. This, however, clashes with the typical teaching of responsable disclosure.

# Table of Contents

## Introduction

From the time I was first introduced to computers I was enamored with the idea of GNU's not Unix (GNU), Berkeley Software Distribution (BSD), Linux, and other forms of free open source software (FOSS). Both the philosophy and the software itself was foundational to my childhood. Naturally, this interest developed into experimenting with and trying out novel forms of software as a hobby. In turn this hobby soon involved Linux distributions in a popular activity called distro-hopping.

"Distro-hopping is a time-honored pastime, where you download and try different Linux distributions."[1]

As someone who's heavily invested in cybersecurity, I often evaluate and experiment with software that can improve my security and privacy. In this way I came across FydeOS, a popular operating system designed to prioritize security and privacy. During the casual inspection of FydeOS, I stumbled across what is often called a "vulnerability", or a hole that would allow a threat to harm a machine running FydeOS. This, for example, could give a threat actor the ability to steal private information and attack other machines. I noticed a compromising and simple security vulnerability that exists in the current version. This vulnerability puts FydeOS's user base at substantial risk.

In this thesis, I will describe the field of cyber security, the role of people like me who look for vulnerabilities, along with the actions taken after a vulnerability is found. This thesis will cover the discovery, the vulnerability, the disclosure process, along with it's further investigation.

## Background

**A Brief Overview of Information Security**

When discussing the security of an operating system it's important to know what you are securing, how you're securing it, and what attacks you might see. A high level overview often includes "models for discussing security issues", "types of attacks", "threats, vulnerabilities, and risk", "risk management", and "incident response".

**The CIA Model**

There are three widely accepted primary concepts of cyber security: confidentiality, integrity, and availability. Each of these hold an important role in keeping a service or product secure. [2]

Confidentiality is concerned with protecting data from unauthorized access and disclosure. It ensures that sensitive information remains private and accessible only to those who have permission to view it. Confidentiality measures can be implemented on many levels of software, from the physical and hardware to the high level software. Examples of multi-layered confidentiality might include a door lock on a server room, hardware or software disk encryption, and the Transportation Layer Security (TLS) encryption that secures the connection between a client's browser and a server.

Integrity focuses on ensuring that information and data are reliable and accurate. It protects data from being tampered with, altered, or falsified by unauthorized actors, entities, or phenomena. Maintaining integrity means that data must not only be protected from deletion but also from unauthorized modifications. Techniques like digital signatures, checksums, hashing algorithms, and version control are used to ensure integrity. Much like confidentiality, integrity can be ensured on multiple levels. On the hardware level, registered error correcting memory (Reg ECC) used in enterprise servers protects data from the corrupting effects of cosmic radiation [3]. At the software level, modern programs are often digitally signed so that they can be recognized by the operating system as secure and intact code, downloaded files verify their integrity through a system called "hasing" to ensure their integrity, and some file systems, like ZFS, have built in error correcting capabilities. Tools like these provide mechanisms to detect changes made without proper authorization and to correct them if necessary, ensuring that data and information stay secure and immutable.

Availability refers to the assurance that authorized users have readable access to information, assets, and services when required. This principle encompasses the ability of systems and networks to remain operational and capable of responding to requests when needed. Threats to availability include natural disasters, power outages, software crashes, and deliberate attacks. Much like all the previous principles, this can be applied on multiple layers. On the hardware layer this might look like avoiding single points of failure, often with the use of redundant systems such as redundant power supplies, multiple servers hosting the same service (high availability), hot swappable hard drive and components, redundant connections to internet and grid power, uninterruptible power supply backups (UPS), as well as the use of the storage techniques like "redundant array of inexpensive disks" often called RAID. On the software side, platforms like kubernetes (k8) and OpenShift allow for software container redundancy, while services like cloudflare protect from denial of service attacks. Availability also includes the client's ability to parse the provided information. For example, a host of spam bots cluttering your favorite social media could be considered a failure of availability as it stops the end users from receiving information the service is hosting.

The CIA triad serves as a framework for understanding and addressing various security challenges. Each element interacts with the others, and compromises in one area can affect the other two. For example, a failure in keeping a password confidential can lead to an attacker changing data and halting services, affecting integrity and availability, respectively. A failure in integrity can lead to an attacker changing a password, which can lead to procurement or dissemination of confidential information and a halt to services, affecting confidentiality and availability.

A failure in availability can lead to data corruption should servers or services fail to shutdown properly; in turn a failure in a security systems availability, such as a firewall appliance, might allow attackers to abscond with confidential information.

**Types of Attacks**

Attacks can generally be classified into four categories, interception, interruption, modification, and fabrication. The line between these categories is often blurred. Many attacks consist of several stages in which each stage can involve a different type of attack. Each of these attacks targets a different concept of the CIA model.

| Concept | Attack Types |
|---|---|
| Confidentiality | Interception |
| Integrity | Interruption<br>Modification<br>Fabrication |
| Availability | Interruption<br>Modification<br>Fabrication |

**Interception attacks** are when unauthorized users access assets that they shouldn't have access too; such as data, applications, environments, or information. This type of attack is more common than most people think. The most obvious example is unauthorized file viewing, but the most common form of this attack might be eavesdropping or looking over someone's shoulder to view their screen. A friend or family member looking through your texts is another common example of an interception attack.

**Interruption attacks** are attacks that cause an asset to become unusable or unavailable. This most directly affects availability but can also affect integrity, as a sudden loss of service can cause the loss or corruption of data. The most commonly talked about interruption attack is a DDoS or distributed denial of service attack. This is when a group of people overwhelm and crash a service, such as a website, with requests. You have probably dealt with an interruption attack when a friend, sibling, parent, or pet unplugged your computer or the router.

**Modification attacks** involve tampering with an asset. This most commonly affects integrity. An infamous example of this is when a hacker changes a user's password. I suffered from a modification attack recently when my cat laid down on my keyboard while I was drafting an email, adding gibberish to it and sending it prematurely. In this way, my cat compromised the integrity of the email by modifying it.

**Fabrication attacks** can be very similar to modification attacks, but rather than modifying an asset that's already present, they generate an asset. This could be generating communications such as emails, database entries such as adding a new user, web traffic, or even generating processes that claim system resources rendering that service unavailable.

    As previously mentioned there are grey areas where an attack might fall into multiple categories. A DDoS attack is both an Interruption attack and a Fabrication attack. It interrupts a service by fabricating threads, taking up more resources than the system has available. An easier-to-understand analogy is my cat causing a DoS attack on my brain. Her meowing

fabricates distractive thoughts which then interrupt my thought processes, effectively denying my brain's availability to work on my thesis.

**Threats, Vulnerabilities, and Risk**

## Threats

A threat is something that has the potential to cause harm. This might be a malicious actor purposefully taking advantage of a vulnerability, or an ignorant user who accidentally accesses improperly secured confidential information. Threats might also include natural disasters such as storms or earthquakes. A threat doesn't need to be malicious in nature.

Common threats include:
- Power loss
- Distributed Denial-of-Service (DDoS) Attack
- Unsecured documents
- Water or fire damage
- Physical access by a threat actor
- Cats running across your keyboa-jkkkkkkkkkkkkkkkktygggttttttttttttttttttttt

## Vulnerability

A vulnerability is a hole or weakness a threat might exploit. Vulnerabilities are often tied closely to context, as a vulnerability on one product or service might not be present on another.

Vulnerabilities might also include physical weakness such as a server room without access control.

Common Vulnerability Exposure Identifier (CVE ID)

A common vulnerability exposure identifier, or CVE ID, is an identifier for a particular vulnerability or exposure in software. It provides a standardized naming system to uniquely identify each vulnerability, which helps organizations track, prioritize, and manage security issues. The Mitre Corperation is responsible for the CVE naming standard. They're a non-profit organization that works with many governments and agencies such as the United State's Department of Homeland Security's Cybersecurity and Infrastructure Security Agency (CICA). We'll explore Mitre's CVE program more in depth later.

**Risk**

Risk is a concept that embodies the presence of both threats and vulnerabilities. Risk might be understood as the threats multiplied by the vulnerabilities. If there are no threats or no vulnerabilities, then associated risk is also zero; this is largely a hypothetical however, as nothing is invulnerable, likewise there are always threats. Risk is a concept we use that describes the present or future probability of one of the CIA Model's principles being compromised. In simpler terms, it describes the possibility of something bad happening.

If either the threats or the vulnerabilities are zero, then risk is also zero, however impossible that scenario might be. Since threats and vulnerabilities are never zero we accept that there will always be some amount of risk. When talking about risk it's important to think about it in it's specific context. The concept of risk is complex and will always be changing and adapting.

**Risk Management**

Risk Management, the act of managing risk, is often viewed as a flow of actions. We start with identifying assets to be protected, identifying threats, assessing vulnerabilities to our protected assets, assessing risk, then taking steps to mitigate that risk. Risk Management is an embodiment of all that comes before as we'll explore further on. This process requires in depth understanding of threats, vulnerabilities, attack types, and the CIA principles

## Identifying Assets

Identifying assets, sometimes called enumeration, sets the foundation for the rest of the risk management process. While the identification of assets to protect might sound like a simple job, it can be deceivingly difficult. It's arguably one of the most important steps in this process. Failure to identify assets to protect could leave open attack surfaces that not only compromise the undefended asset, but act as a pivot point for attackers to launch further attacks off of.

An example of an often overlooked asset is phone infrastructure like VoIP or PBX. VoIP and PBX are both common methods of providing telecommunications infrastructure to enterprises. The mid 2010's were wrought with hackers who'd hack into a corporation's telecommunications infrastructure. After gaining control they'd call a pay per minute line they'd own. Over a weekend, when all employees are out of the office, a hacker could use every connected phone in the building to call their line, racking up charges in the tens of thousands of dollars.

Let's carry this example over to the rest of our flow of actions. Keep in mind that this is a generalization and, due to brevity, shouldn't be taken as a comprehensive telecommunications risk assessment.

**Identifying Threats**

This is the "what if" stage. Here, we combine our assets with the CIA model to help identify potential threats. For example: We've identified an asset we want to protect, in this case our company's phone lines. Using the CIA model we can dissect threats to our telecommunications.

Which threats affect confidentiality? Wiretapping and compromised PBX servers are both examples of threats which affect the confidentiality of a phone line.

What threatens integrity? A compromised PBX or router might redirect incoming calls to a third party. In such an example, an attacker might be able to impersonate an employee.

Lastly we look at what threatens availability. Call spamming or call flooding can render a business number inoperable as easy as cutting your businesses landline can. Cutting power to the whole building could cause the PBX server to shutdown.

**Assessing Vulnerabilities**

Vulnerabilities must be assessed in the context of a threat. Assets generally have hundreds or thousands of threats, but only a fraction of these will be relevant. Here we explore which of these threats pose a risk due to a vulnerability.

Confidentiality: Wiretapping isn't a risk since our hypothetical businesses phone lines are underground. We keep our network secure, follow best practices for network security to protect our PBX server. As such, there are no known vulnerabilities here.

Integrity: As previously mentioned our PBX server is security. There are no vulnerabilities to integrity.

Availability:We don't have any protection against call spamming. We also don't have any backup power supplies in place for our PBX server. This represents two vulnerabilities to availability.

**Assessing Risk**

Here we match our known threats to our known vulnerabilities. In this case there are two threats. If we have a power loss (threat), and the PBX box is affected by that power loss (vulnerability), then we have a risk of our telecommunication systems failing due to power loss. If a threat actor spams our phone lines with calls (threat), and we have no way to set up phone number filtering, spam protection, or sanitation (vulnerability), then we have a risk of a call bomber causing our phone infrastructure to grind to halt.

**Mitigating Risk**

Risk mitigation can either be done by removing threats or vulnerabilities. Both actions heavily rely on context. In our case, we can remove the threat of power loss by putting an uninterruptible power supply between our PBX server and grid power. This way, in case of power loss, the server will still operate. We've successfully addressed the vulnerability.

We can mitigate the risk of call bombers by choosing a PBX or a provider which offers features such as phone number sanitation, spam filtering, or number whitelisting. By using these features we can mitigate the risk of call bombers.

It's important to note that we are mitigating risk, not removing it entirely. There will always be risk.

**Who Finds Vulnerabilities**

Vulnerabilities are searched for and found by a wide range of individuals and organizations. Before getting too deep into classifying these entities, we should cover some definitions.

Penetration testing, often shortened to pentesting, is the act of actively testing a defense, to identify and exploit vulnerabilities so that they may be patched or fixed later. The purpose of this is to simulate an attack and mitigate risk. For example, a bank might hire a security firm to run a penetration test on their network. In such a scenario, the firm would actively try to hack the bank, then report its findings back to the bank. Pentesting, despite consisting of offensive action, is part of the defensive security framework.

Anyone can search and find a vulnerability, but those that do so professionally can usually be categorized into several categories: independent, organized, and government with motivations ranging from defensive penetration testing, neutral, to offensive actions. Viewing these categories as a three by three matrix can help us understand the players. In the example below, we'll fill in the matrix with a matching threat actor.

|  |  | Organization Type | | |
|---|---|---|---|---|
| Motivation |  | Individual | Organizational | Government |
|  | Defensive | Freelance pen tester and independent security contractors | Private cyber security firm | FBI/CISA |
|  | Neutral | Gray hat hacker (e.g. bored teenager hacks into NASA) | Hobbyist Hacking groups (e.g. Cult of the Dead Cow) |  |

| | Offensive | Individual cyber criminals and hacktivist (Kevin Mitnick) | Organized Cyber Crime and Hacktivist Groups (e.g. Anonymous) | NSA (USA), Cozy Bear (Russia) |

**A Brief Overview of the Penetration Testing Pipeline**

According to The Penetration Testing Execution Standard, there are seven steps to avery engagement. [4] These steps are: Pre-engagement Interactions, Intelligence Gathering, Threat Modeling, Vulnerability Analysis, Exploitation, Post Exploitation, and Reporting.

The pre-engagement stage acts as a starting point for all the parties to communicate. In this stage a client will define a scope and rules of engagement. A scope is a range of targets which the penetration tester is allowed to target. The process of a penetration test can, even inadvertently, be destructive. A client might also not want to expose their data. As such, a scope of allowable targets is defined. Often a client might set up a mirror, or a copy of their production servers. This way they can test their security model with minimal exposure. Also included in this stage is pay and timelines.

The next step is intelligence gathering. Sometimes called reconnaissance, the intelligence gathering stage involves several sub steps. Targets during this phase not only include digital assets but might also include corporate/bureaucracy structure, finances, researching employees, and scouting physical locations. Reconnaissance targets may also include individuals, such as researching an employee's background, looking for court records, political stance, social media, and digital footprint. Human intelligence gathering (HUMINT) is a very powerful tool. HUMINT might involve the penetration tester assuming a fake identity and befriending target employees. If in scope, a penetration tester might even scout physical locations.

Threat modeling is the process in which information is categorized, organized, and analyzed to better understand the threat landscape. Here we identify and categorize priority and secondary assets, identify and categorize threats, then finally we map each threat to each asset. When finished we should have a comprehensive understanding of the threat landscape. It's important to remember to include physical security and humans in your threat model.

Vulnerability analysis is the process of discovering vulnerabilities that can be leveraged by an attacker. Vulnerability analysis can be broken down into three areas: active, passive, and research. Active analysis involves actively testing services, applications, and networks to find vulnerabilities. Passive analysis involves activities such as metadata extraction and network scanning which might reveal a vulnerability (e.g. a publicly available pdf's metadata contains an internal server address). Finally research involves checking services and applications against known CVEs and vulnerabilities.

The exploitation stage involves using that vulnerability to establish a connection with a system. Here a penetration tester might craft an exploit and malware payload. They might take into consideration anti-virus software, obfuscation techniques, and physical access. Depending on context, the penetration tester might analyze publicly available source code, fake API requests, or perform code injection.

Once a beachhead is established, point exploitation can begin. Here the penetration tester might attempt to escalate permissions, gather confidential information (sometimes called pillaging and the collected information loot), or perform further penetration attacks on other systems. Common goals during the post exploitation stage involve gaining persistent access, pivoting to attacking other systems, and generally making it difficult for an incident response team to remove their access.

The last stage is reporting. Here the penetration tester crafts a report with their findings. This will often include a background, overall security assessment rating the difficulty of access, recommendations to fix any vulnerabilities, a strategic roadmap, and a technical report.

**What is FydeOS**

FydeOS and its sister operating system, OpenFyde, are web-centric, operating systems based on ChromiumOS meant to be a drop-in replacement for ChromeOS. Web centric in this context means that, rather than user interactions being focused on the local environment, the operating system is optimized for the most interactions to be on the web. Rather than using local applications like Microsoft Word, users are expected to use cloud services like Google docs.

FydeOS is developed by the company Fyde Innovations based out of Haidian District, Beijing. It's a popular Chromium-based operating system designed to prioritize security and privacy; it does this by

FydeOS's website specifically mentioned: Simple, Swift, and Secure on their front page.[5] A user who doesn't have a background in cybersecurity would have little to no way to verify this claim of security.

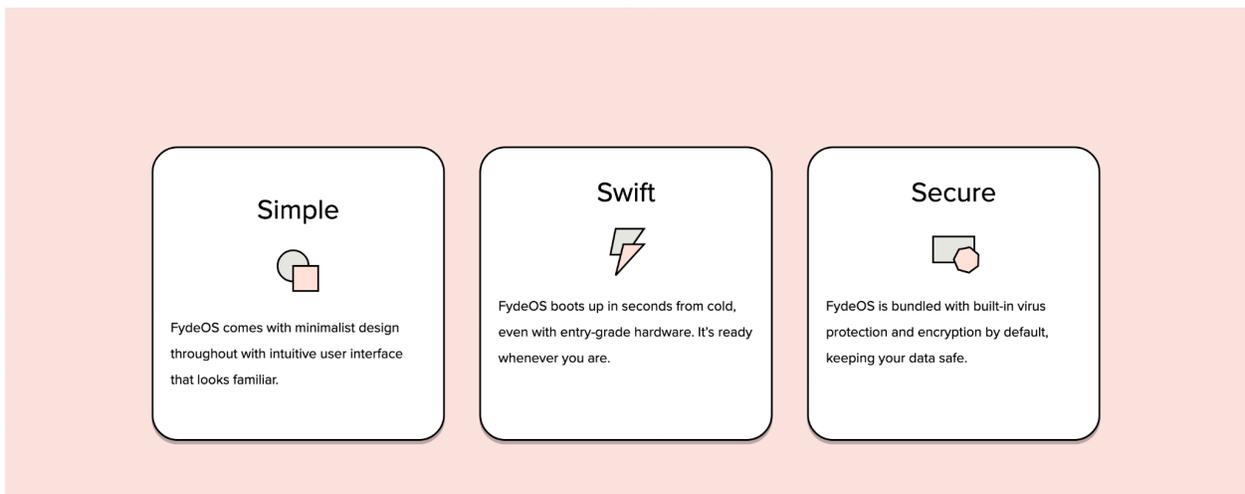

FydeOS has several pages on their website talking about security in their FAQ section. [6], [7]

"Is FydeOS secure?

Absolutely.

> FydeOS is engineered with security as a priority, making it a reliable choice for your daily computing needs. Unlike traditional Windows PCs, FydeOS eliminates the necessity for anti-virus software or frequent malware scans, offering a seamless and worry-free experience."

> "Does FydeOS carry malware that might compromise my information security?
>
> We can only guarantee that operating system images distributed from official FydeOS download page bearing the FydeOS logo are secure, harmless and comply with our Privacy Policy and other legal provisions relating to Data Protection Act in the United Kingdom."

We'll examine these quotes in greater depth later on.

**FydeOS vs OpenFyde**

FydeOS and OpenFyde are siblings. They share much of their code base and have a nearly identical user interface. Their largest differentiating factor is their licensing. OpenFyde, as its name might suggest, is an open source distribution of Fyde. It acts as an upstream and base for FydeOS. FydeOS is a downstream distribution of OpenFyde with added proprietary features that streamline the user experience.

**Discovery**

I first came across FydeOS while looking for an alternative to ChromeOS. ChromeOS has great usability, is lightweight, secure, and robust. All good features. Because it's light weight it's a great option to revive a poor performing device and give it a second life, and because it's secure, easy to use, and robust it's perfect to give to a family member or friend.

Unfortunately ChromeOS also is closely tied with Google and tracks both your data and you. While great for security and usability, it's extremely bad at safeguarding user privacy. FydeOS promised to provide a ChromeOS experience, with its usability, security, while also removing the Google aspect, promising privacy. On the surface it seemed like a great alternative.

As part of my standard operating procedures, when exploring new software and technologies I plan to use for personal use, I often perform a quick and dirty security audit. If I plan on putting sensitive or personal information on this device I want to make sure it's reasonably secure. The purpose of this audit isn't to scan and root out any major vulnerabilities, but more to gauge the developer's priority on security and privacy. Have the developers taken steps that show that security is a priority for them? Are there any obvious security flaws and have they taken any shortcuts which would cause me to doubt the security of the rest of the system?

This process is a bit like kicking the tire of a car before buying it. You don't expect such a simple test to tell you much about the vehicle, but if it's low on pressure you might question it's quality of care and maintenance. You definitely wouldn't expect the car to fall apart. Unfortunately, that's what happened during my inspection of FydeOS.

**Users and Groups**

Users and Groups are the basis for access control on Unix systems. A user is typically a person or entity that uses a system. In the context of Unix, a user represents an account or digital identity associated with a person, entity, or process that's allowed specific access to a system. Each user has a unique identifier, a number, that's associated with their account. This user identifier is called a UID and acts as a numerical representation of a user of a program. Some UID and UID ranges are reserved for specific purposes.

| UID | Purpose |
| --- | --- |
| 0 | reserved for the root or administrative user |
| 1-99 | System daemon accounts |
| 100-999 | User daemon accounts |
| >=1000 | Standard Users |

As you may have noticed, only UIDs after 999 are used for standard users. Many background processes, often called daemons, and programs are given their own user account. This allows processes to compartmentalize their permissions which protects them from interfering with the greater system as a whole.

Groups, as their name suggests, are subsets of the total user set (or groups of users). These allow for grouped access control. For example, on Unix-like systems, there is often an administrator group called Sudo or Wheel. These groups are given elevated privileges and can run commands normal users usually cannot. Users can be added and removed from groups. This increases the ease with which access control can be configured, much the way a corporation might group employees into departments and compartmentalize access to parts of a building. Only those in finance can access some areas of the finance floor, and if an employee is moved from finance to sales, they need merely switch their group rather than rekey every door in finance and sales to accommodate the change. Back to the computer, each user is part of a primary group. This is usually their own, individual group, matching their username. Each user can be assigned to any number of groups on top of their primary group.

For example, a standard user named `foobar` might be a part of groups `foobar, sys, log`. This configuration would allow user Foobar access to their own group files, printers, and log files respectively. Likewise, their access can be revoked by removing them from a group. Like users, groups have numerical identifiers as well, called GID.

Users and Groups are the building blocks of Unix security and access control. While we'll be focusing mostly on users, it's important to have a holistic picture of Unix's access control schema.

**Teletype - TTY and PTY**

The concepts of teletype (TTY) and pseudo-teletype (PTY) in computer science date back to the early days of computing and the dawn of Unix. Teletype machines, formally called teleprinters, were used for decades before their induction into the world of computers. They were used to send and receive typed messages. A mix between a typewriter, a fax machine, and a telegram, teletype machines could both send and receive messages while keeping a local copy on a paper sheet. As the user typed, their message would both be recorded on a local paper sheet, like a normal typewriter, while also being sent out of the machine with electronic signals. The machine would record incoming messages on that same sheet of paper. The resulting sheet of paper would hold a transcript between the user and the recipient.

In the days of early computing, before we had screens, teletype machines were one of the best ways to interface with a computer, and the only way to do so remotely. By the 1960s, Teletypes were widely used as terminals for remote access to mainframe computers. Pseudo-terminals were introduced later. These enabled multiple teletype sessions on a single physical terminal. To this day, the concepts of TTY and PTY are not only still used, but essential components of Linux systems.

On modern computers the functionality of a teletype machine is simulated with the use of a virtual console. The TTY's functionality under the hood has largely stayed the same. To this day we identify terminals and consoles with the prefix TTY or PTY then a numeral. For example TTY2 or PTY4. To switch from one TTY to another Linux users can often press the keys `control+alternative+fn`. For example, to switch to TTY2 a Linux user might press

`ctl+alt+f2`. On many more locked down systems, such as ChromeOS, this feature is disabled.

**What is Root?**

The root user is a special and unique user in a system. Given the UID and GID "0", the root user can change the ownership and permissions of any file. Named after the user who owns the root directory, the base directory on a system (abbreviated as "/"), the root user has full control over the system. The Microsoft Windows equivalent is the Administrator user account. This account is a high-priority target for any hacker or penetration tester. Due to the privileges given to the root user, if it's compromised, it compromises the entire system. Confidentiality, integrity, and availability can all be controlled by the root user.

"Root privileges are the powers that the root account has on the system. The root account is the most privileged on the system and has absolute power over it (i.e., complete access to all files and commands). Among root's powers are the ability to modify the system in any way desired and to grant and revoke access permissions (i.e., the ability to read, modify and execute specific files and directories) for other users, including any of those that are by default reserved for root." ~ The Linux Information Project [8]

"If you lose access to the root account, the person who gains control literally becomes God and you have effectively lost control of your system." ~ Hunter O'Gold [9]

**Configuration - What It's For, Where It's Stored**

The programs that make up an operating system need ways to change dynamically with different situations. Some common examples of this might be changing a desktop background, adding a user, changing a password, joining a wifi network, or even changing when your computer decides it should sleep. These values change the behavior of the program referencing them. Collectively these values are called a configuration. These are often kept in plain text, easily readable and modifiable, and are kept in a file. Programs will then reference these files and behave accordingly. For example, the name of a system, often called its hostname, is kept in the file /etc/hostname. Generally, when a Linux operating system starts, a program called systemd will perform a system call, a low-level function, and set the hostname to whatever text is in the /etc/hostname file. By changing the text found in the configuration file, you can change the behavior exhibited by the system, in this case, the system's name.

Configuration files change the behavior of the system. Examples include the definitions of users, groups, and permissions, and the configuration of services such as secure shell (SSH), which allows users to log in to a system remotely. You've probably interacted with configurations before without even realizing it. When connecting your device to a wifi network, changing the power profile on a laptop, or putting your phone on do not disturb, you're changing its configuration. The configuration options available in a device's graphical user interface are typically rather innocuous. Operating system designers will avoid putting volatile or unsafe configuration settings where they can easily be changed by users, but that doesn't mean that these unsafe configuration files don't exist.

When searching for a vulnerability we often examine the way software behaves, and by extension, its configuration files. By knowing a program's behavior we gain a better understanding of the operating system as a whole, and in turn, a better understanding of where to look for vulnerabilities. In a study conducted by Open Worldwide Application Security Project (OWASP), in 2021, 19.84% of all applications tested were vulnerable due to misconfigured security configurations. [10]

Part of my standard operating procedure when trying a new software is ensuring it is secure. In the case of operating systems, this means reviewing the configuration files, among other steps. According to the Filesystem Hierarchy Standard, a (FHS)[11], "The /etc hierarchy contains configuration files", and is thus our first stop.

The /etc/ directory has a history dating back to the first initial Unix v1 release in 1971. [12] Standing for "etcetera", the directory was meant to hold everything that didn't have a home elsewhere in the system. It soon became the de facto home for system configuration files as, at the time, those didn't have a dedicated home. When the FHS was first released in 1994, the ad-hoc solution became an official standard. As such most system-wide configuration files can be found in the /etc/ directory of most Unix-like operating systems.

**Password Authentication on Unix-like Systems**

Usernames and passwords are a common access control model. The implementation of such a model involves a username, or user identifier, and an authentication factor, such as a password or passphrase. This model is often used by the Pluggable Authentication Module (PAM) on many Unix-like systems.

Usernames are public identifiers and are not typically sensitive information. Passwords on the other hand are sensitive information as they're what allow usernames to be authenticated. As such they must be protected.

Passwords should be long and complex enough not to be easily guessed. They should also be encrypted so that they're not compromised if the file containing them is leaked.

On modern Unix-like systems, passwords are encrypted using a hashing algorithm and then stored in a file. You can think of a hash algorithm as a cryptographic one-way valve. In this case, you put secure data, such as a password, in one end of the function, and you get seemingly random data out the other end. What's important is that if you put the same data in, you get the same data out. It's extremely difficult or impossible to reconstruct secure data from the output.

Example:

A popular hashing algorithm is md5. Let's pretend our password is mypassword. The corresponding hash for this password is 34819d7beeabb9260a5c854bc85b3e44. We call the output "the hash" of whatever the input data is. In this case 348…e44 is the hash of mypassword.

Hashes are vital to the authentication architecture of modern computers as they allow the system to compare a given password with a user's password without ever storing that password.

When you set up a user account on a unix-like system and first set your password, the computer will compute your password's hash, and then store that hash in a special file in the /etc/ directory. When you later log in, your password is again hashed and then compared to the original, stored, hash. If the two hashes are the same, then the system assumes the passwords are the same and allows the login. Passwords are stored in hashes rather than plain text so that if the password file ever leaks, the user's password is still encrypted.

**Anatomy of the /etc/passwd and /etc/shadow file**

While /etc/ holds many configuration files we are going to focus on two, the first of which is the /etc/passwd file. This file is responsible for defining users and holds user account information. The second is /etc/shadow which holds "shadowed" user information, such as password hashes. Shadowed information refers to information which should be kept confidential. Most of the files found in the /etc/ directory are encoded in ASCII, one of the first binary to text standards. This makes them easy to read and edit by both humans and programs. These two files, /etc/passwd and /etc/shadow, act as a record of users and their associated system level configurations. This is separate from the user's personal configurations, such as desktop backgrounds.

/etc/passwd is responsible for storing user account information. Each line is responsible for a user and each field is delineated with colons. [13] The fields are as follows:

| Field Name | Username | Password | UID | Primary Group GID | Comments | Path to home directory | Path to default shell |
|---|---|---|---|---|---|---|---|
| Example | root | x | 0 | 0 | Root User | /root/ | /bin/bash |

When in the proper format this information would look like the following entry:

```
root:x:0:0:Root User:/root:/bin/bash
```

Historically, the password hash used to be stored in the /etc/passwd file. Since this file needs to be referenced by many programs, its permissions are looser than is proper for safely storing password hashes. To remedy this, password hashes were moved to a separate file, /etc/shadow, with tighter permissions. The character x is used to tell programs reading the /etc/passwd file that this user's associated password is kept in the /etc/shadow file.

Like the /etc/passwd file, the /etc/shadow file is written in plaintext with an entry on each line. Its syntax though is a slightly different format. [14]

| Field Name | Username | Password | Last | Min | Max | Warn | Inactive | Expire |
|---|---|---|---|---|---|---|---|---|
| Example | foobar | [example below] | 19770 | 0 | 99999 | 7 | 30 | 0 |

The Username field, as previously described, acts as an identifier for the user's account.

The Password field stores the user's password information. We'll go into greater detail about the password field below.

The Last field records when the user's last password change was. The date is recorded in Unix time (sometimes called POSIX time), a special format in which a date is expressed by the number of days since Jan 1st 1970. The value 0 is reserved to signify that the user must give a new password upon login.

The Min field records the minimum number of days between password changes. In other words the number left before a user may change their password again.

The Max field expresses the number of days the given password is valid for before being forced to change their password again. In other words, this field contains the number of days left before the password expires and the user must pick a new one.

The Warn field expressed the number of days before their password expires.

The Inactive field expresses the number of days an account remains active after a password expires. For example, if this value was set to 7, if a user were to not log in for seven days after their password expired, their account would be locked.

The Expire field holds the date, in Unix time, in which the account was first made inactive.

The password field, which we've skipped over, is a bit more complex. It's made up of three subfields, a hash ID, a salt, and a hash. Each of these fields starts with a $. [15]

Anatomy of the Password Field in the /etc/shadow file:

**Password Field Syntax**

| Password Field | Hash ID | Salt | Hash |
| --- | --- | --- | --- |
| Example | $1 | $salt | $hashd653ea7ea31e77b41041e7e3d32e3e4a |

The hash ID signifies which hash algorithm that user is using.

The salt is composed of random characters which help add randomness to the hash. Using a well randomized salt we can ensure that the same password's hash will look different on two different machines.

Finally we have the stored hash. As we've covered in the previous section, the hash is an encrypted version of the password.

Wild cards are a symbol, character, or sequence of characters (often an asterisk "`*`") that's used to represent the set of any combination of characters or none at all. For example, given a set of integers between the numbers 0 and 100; a search on that data using a wildcard would return the entire set, including the empty set. A search using `2*` would return every number starting with the number 2, in this case 20-29. A search using `*0` would return every number ending in zero; in this case 0, 10, 20, … 90, 100.

**The Fyde password vulnerability**

Now that we understand concepts like root users, password authentication, and the importance of configuration files, let's take a look at FydeOS's /etc/shadow file.

```
localhost / # cat /etc/shadow
chronos:*:::::::
root:*:::::::
```

Ohhh… Ohh no. FydeOS seems to have left the password for the most important user on their operating system effectively blank!

Unfortunately it gets worse. While ChromeOS, the basis for FydeOS, removed the ability to switch TTYs, FydeOS kept that functionality. This means that, without even logging into or getting past the first screen, an attacker can press `ctrl+alt+f2`. When the console login screen prompts them for a username they need only enter the word `root` and they gain full access to the whole system. FydeOS has a wild card set as its root password.

This is what we call a "default password vulnerability". This can be an incredibly damaging vulnerability on non privileged users, on a root account this is catastrophic.

     A concern with this particular vulnerability is how blatant it is. Configuration files have to be written by someone. This means that someone, at some point, chose for the root password to be a wildcard; this is not something someone does by accident. At this point in our story, whether this was done by FydeOS or an inherited feature, such as from a dependency, is unknown.

This type of vulnerability threats every part of the CIA model. Confidential information, such as files and saved passwords stored on the system might be extracted. Files might be changed and manipulated. Lastly, the system could be completely deleted.

**Disclosure**

**Who Uses FydeOS - Users, Clients, and Exposure**

Part of assessing the exposure a vulnerability presents is by looking at affected parties. Unfortunately, FydeOS has a large client list which include:

- Public libraries in Europe and the UK
- Brazilian medical technology firm
- Plastic Packaging Germany
- Scrap Metal Recycling Company
- European Agricultural Enterprise
- Industrial Component Manufacturing Organisation

How much exposure does this vulnerability have? "FydeOS - a proven Chromium OS distro that has been maintained for over 6 years and has millions of users worldwide." [16] That's potentially millions of exposed users.

**fydeos.io vs fydeos.com**

Fyde Innovations has two subsidiaries, each of which has a website to distribute their own version of FydeOS. The first subsidiary is headquartered in the United Kingdom; their website domain is fydeos.io. The second subsidiary is based in Haidian District, Beijing in China, where Fyde Innovations is located. For the remainder of this paper we'll refer to them as io and com. The com domain is associated with China and the Chinese market, while the io domain is associated with the United Kingdom and the global market.

**Nation State Actors Considerations**

In the context of cybersecurity, nation-state actors refer to nation-states or state-sponsored entities that use cyber capabilities to achieve strategic objectives. Nation-state actors can be both offensive threats and defensive.

Offensively, nation-state actors are often seen engaging in activities such as denial of service, espionage, and disinformation campaigns to gain an advantage over their adversaries. For example, the United State's 2016 Democratic National Committee (DNC) hack is widely attributed to the Russian state sponsored groups "Cozy Bear" and "Fancy Bear". It's believed that these nation state actors sought to influence the US presidential election [17].

Nation-state actors can also play a crucial role in defending their countries' cyber interests. They may invest in cybersecurity research and development, establishing capabilities to detect and respond to threats. For instance, the United States has established its Cybersecurity and Infrastructure Security Agency to coordinate the country's cyber defense efforts, while China has set up its National Computer Network Emergency Response Technical Team (CNCERT) to monitor and respond to cyber threats.

**Backdoors**

In cybersecurity, backdoors refer to hidden persistent access an attacker leaves behind after compromising a system. This allows them to maintain persistent access over a system without detection, effectively creating a secondary hidden vulnerability. Backdoors grant attackers greater flexibility during cyber operations, enabling them to:

- Separate the initial exploitation of a vulnerability from the delivery of a payload or an attack.
- Re-infect a system if an incident response team clears out the malware.
- Selectively target compromised systems, minimizing information overload and traceable activity.
- Continuously monitor and control the system.

Backdoors are one of the key tools used by Advanced Persistent Threats (APTs). APTs use backdoors to evade detection, dynamically modify systems, engage in multi-stage attacks, and perform lateral movements through a network. APTs can infiltrate a network and stay undetected for years; even when caught, they're difficult to completely root out and might hibernate or go dormant for months only to resurface later. They're one of the most dangerous types of threats.

By exploiting these backdoors, attackers, like APTs, can prolong their access to a system, making it challenging for defenders to detect and respond effectively. They are commonly used to establish a command and control (C2) channel, conduct reconnaissance, and perform lateral movements.

**Poisoned Code**

A popular tactic employed by nation state actors is that of poisoned code. Poisoned code is code that is intentionally made vulnerable or contains malware that is injected into a legitimate program or application.

A recent example of poisoned code is the XZ Utils Backdoor. XZ Util is a lossless compression utility tool for Unix-like systems. On March 29th 2024, Andres Freund found an obfuscated back door hidden in the upstream binary. [18] Registered as CVE-2024-3094, this backdoor would have had one of the largest exposure surfaces of any poisoned code in history. As nearly every distribution of Linux and MacOSX uses XZ Utils this could have had extremely serious ramifications. It's not hyperbole to say that if the backdoor had a chance to propagate as intended, the threat actor would have had the power to change human history. Alex Stamos, former chief security officer at Facebook was reported saying "This could have been the most widespread and effective backdoor ever planted in any software product," [19] Fortunately it was caught before many systems could update to the poisoned code. The story of the XZ Utils Backdoor is complex, sordid, and still developing. As digital infrastructure grows in complexity and dependencies, so does their attackable surface area.

**Poisoned Fyde**

FydeOS is in a unique position for an operating system. Notably, it's based in China, while the majority of its user base is outside of China. With governments and corporations all around the world using FydeOS, the possibility of this vulnerability being intentionally poisoned code by a state sponsored actor shouldn't be ignored. What makes this particular vulnerability particularly odd is how obvious it is. This is a very amateur mistake that should have never gotten past the prototyping phase. Despite this obvious vulnerability, FydeOS advertises the security of their operating system. These two things are incongruous.

While the likelihood of the vulnerability found in FydeOS being sponsored by a nation-state actor is low, it still warranted further investigation. Of particular concern is the fact that the Chinese version of FydeOS and the global version of FydeOS are different files, as we'll explore later.

**Disclosure Types**

When a vulnerability is found there comes a difficult question. What do you do with that information? What's the best way to safely and ethically disclose what you've learned? Disclosure refers to the action to be taken after finding a vulnerability. There are three primary forms of disclosure. Private Disclosure, Full Disclosure, and Responsible Disclosure.

Private Disclosure is when the vulnerability is reported privately to the parent organization. It's up to the organization to decide if they want to publicly disclose the vulnerability and how they want to address mitigation, fixing the vulnerability. This process allows organizations to silently address a vulnerability without the public ever knowing it was there. The pros of this approach is that bad actors might never learn of the vulnerability and will have no chance to exploit it, the cons is that if a bad actor has already taken advantage of a vulnerability, the end users might not be aware. Unfortunately, not all organizations take security seriously. In cases such as this, they may not want to fix the vulnerability or take any action to mitigate the issue. In such cases, some people resort to Full Disclosure.

Full Disclosure is the most controversial form of disclosure. It's most commonly used as a response to an organization's failure to respond properly to private disclosure. Full Disclosure dictates that you publicize the details of the vulnerability. This pushes parent corporations to not ignore a vulnerability and allows their respective communities to put pressure on the parent organization to fix it.

Responsible Disclosure, sometimes called Coordinated Disclosure, is a middle-ground approach between private and full disclosure. In this method, there's a private initial report, often to a few trusted organizations, so that the company has time to address the vulnerability. After a set time, or after a patch has been released and has had time to propagate, the vulnerability is fully disclosed. This is the default method for most independent security researchers today. A researcher will often give an organization a deadline to produce a patch; usually between 30 and 90 days. If the organization doesn't publish a patch, the researcher might publish it publicly after this time.

| Disclosure Type | Audience | Pros | Cons |
| --- | --- | --- | --- |
| Private/Closed | Parent Organization | Reduces the chance of threat actors using the vulnerability | Organizations are not pressured to mitigate the vulnerability |
| Public/Full | Public | Organizations are | Threat actors can use |

| | | pressured to take quick action to address the vulnerability | the vulnerability before a patch is published and integrated |
| --- | --- | --- | --- |
| Responsible/Coordinated | Trusted Organizations, then Public | Reduces the chance of threat actors using a vulnerability, but still pressures organizations to patch | Threat actors have a small, but still real chance to use the vulnerability before it's patched. |

The proper disclosure of a vulnerability is vital to being an ethical security researcher. If the vulnerability isn't disclosed properly it can lead to threat actors using it before the vulnerability can be patched, or it may never be patched at all. As such, security researchers use techniques like responsible disclosure to ensure that security issues get addressed while reducing the number of entities at risk.

An example of failure in responsible disclosure was the Eternal Blue exploit.
Eternal Blue is an exploit that takes advantage of CVE-2017-0144 [20]. The exploit was developed by the United States of America's National Security Agency (NSA) around 2015. In 2017 the exploit was stolen from the NSA and auctioned by a group called The Shadow Brokers. From there it went on to be one of the world's worst computer viruses, being the basis for the infamous WannaCry ransomware.

Responsible disclosure is a concept that aims to stop situations like that surrounding the release of Eternal Blue from happening again. As part of responsible disclosure, security researchers can report their findings to several organizations.

**FydeOS Disclosure**

For the disclosure of this vulnerability I selected three entities. The first is Fyde Innovations, the developers of FydeOS. Their help is needed to not only fix the vulnerability in future releases, but to inform their current users about the issue. The second entity to inform is the United State's Cybersecurity and Infrastructure Security Agency (CISA). Formerly, United States Computer Emergency Readiness Team (US-CERT), CISA is a department of Homeland Security and, as their name suggests, is responsible for cybersecurity and infrastructure. Any vulnerability that affects the US government or its infrastructure is within its purview. The last entity to disclose this vulnerability to will be Mitre. Mitre (pronounced mite-ter) is a non-profit, independent national security advisory. Dan Macsai, the former executive editor, and current chief events officer, at Time Magazine, once called Mitre "The most important company you've never heard of".[21] Mitre run the CVE Program whose aim it is to "identify, define, and catalog publicly disclosed cybersecurity vulnerabilities" [22].

**Fyde Innovations Response**

As part of responsible disclosure, we must inform the parent company. In this case FydeOS. As I was already a member of the FydeOS Discord Server I asked about bug bounty programs on their support channel. An employee at FydeOS, who I later learned was it's co-founder and CEO, responded saying:

@fydeos_alpha

> FydeOS for PC / You are shipped in dev mode, ie, not air tight and welcomes hacking. FydeOS in the enterprise solution has a totally different approach. We don't have a bounty program, but if you are willing to help us/other rather than exploit vulnerabilities, you can let us know about this by email (edited)"

> @fydeos_alpha
>
> It's true that we have left the shell access open for all public releases of FydeOS, including FydeOS for PC, for You and all openFyde builds. This is the default behaviour for dev images that are configured by the Chromium OS SDK. We could talk more on this later.

> @fydeos_alpha
>
> Admittedly, publicly available FydeOS releases aren't as secure as ChromeOS on Chromebooks on a default setting, but it should behave the same when a chromebook turned into dev mode. I agree that we have not publicly emphasis the difference between FydeOS and default ChromeOS in terms security, this is something we will definitely look into and improve.

After a discussion, it became clear that FydeOS was aware of some of the security issues but none of this information was ever disclosed to the end user. Not only did they knowingly release an operating system with security issues, they told their user base that it was secure; while keeping the secure versions for their paying customers.

This behavior raises several ethical concerns. First and foremost is the lack of transparency. FydeOS didn't disclose the implications dev mode has on user security, despite being aware of it themselves. This lack of transparency is a clear violation of users' trust. This is only made worse when you take into account the amount of marketing material FydeOS has boasting of their operating systems security. These actions, when taken together, might be seen as a fragrant disregard for their users security and privacy.

Second is informed consent. Users have the right to informed consent regarding the security vulnerabilities and the potential risks associated with using an operating system. FydeOS failed to provide this essential information, leaving users vulnerable and unaware of the risks.

Finally FydeOS response, when I brought up a security issue, was to say that their secure operating systems encourages exploitation. To me, this communicated that, when paying customers weren't involved, they didn't value their operating systems security. This behavior not only compromises user security but also undermines the trust that users have in FydeOS and its commitment to security.

To their credit, FydeOS did accept that they needed to better communicate with their users. "I agree that we have not publicly emphasis the difference between FydeOS and default ChromeOS in terms security, this is something we will definitely look into and improve." We can only hope that it's not too little too late.

While Fyde's response was disappointing, this behavior isn't wholly unusual for open source projects. It's very possible that Fyde's running on thin margins, has a small development team, and might not have the labor needed to properly secure their operating system. Many open source projects forgo security to achieve a minimum viable product. This, in itself, isn't wholly bad, so long as this is disclosed. What makes Fyde's particular situation disappointing, however, is that they misled, if not lied, to their user base about the security of FydeOS. I blame them for the lack of security, but the lack of communication.

**CISA / US-CERT**

Most countries will have a Computer Emergency Readiness Team (CERT), or a team, department, or agency that fills the same role. These teams are often composed of cyber security professionals that are trained to quickly perform incident response and vulnerability mitigation. They'll often have a hotline phone number by which you can quickly reach them. These teams are the 911 or emergency dispatch services of the cyber world. Some examples of of these teams are:

1. The United States Department of Homeland Security's Cybersecurity and Infrastructure Security Agency (CISA)
2. The European Union Agency for Cybersecurity (ENISA)
3. The Japanese Information-technology Promotion Agency (IPA)
4. The Israeli National CERT (INFCERT)
5. The Australian Computer Emergency Response Team (ACERT)
6. The Indian Computer Emergency Response Team (CERT-In)
7. The Canadian Centre for Cyber Security (CCCS)

8. The United Kingdom's National Cyber Security Centre (NCSC)

9. The French Agency for the Security of Information Systems (ANSSI)

10. The German Federal Office for Information Security (BSI)

What's considered "proper action" might change depending on the affected parties, your location, and your citizenship. In my case, it proper action consisted of reaching out to the United State's Cybersecurity and Infrastructure Security Agency (CISA)

Disclosure to CISA was rather straightforward. I found their phone number online, called it, and communicated my findings.

**CVE Registration**

The last entity that I've disclosed this vulnerability to is Mitre and their CVE Program. Mitre's Common Vulnerabilities and Exposures (CVE) program is an internationally recognized standard for identifying and classifying security vulnerabilities in software, hardware, and firmware. Launched in 1999 by the Mitre Corporation, the CVE program was initially designed to provide a centralized system for tracking and naming vulnerabilities, allowing developers, users, and security professionals to efficiently communicate about and address potential weaknesses. [22], [23]

The CVE program has, over years, evolved to become an essential component of the global cybersecurity landscape. The CVE Identifiers (CVE IDs) are unique strings that identify a specific vulnerability, providing a common language for describing and tracking vulnerabilities across different products and platforms. By standardizing the naming convention for vulnerabilities, the CVE program allows for clearer and more accurate communication and sharing of knowledge between security researchers, developers, and organizations worldwide. When a security researcher finds a vulnerability, it's typical that they'd want to register their find. There are many reasons to register a CVE. Many security researchers do so as public service, some register just to add another feather in their cap. Like World War II victory markings on fighter planes, CVEs can act as a record of a security researcher's accomplishments, providing both bragging rights and padding to one's resume.

The process of registering a CVE is formally known as "Requesting a CVE ID". In this process a security researcher will submit a form containing the details of the CVE to a Certified Numbering Authority (CNA).

"CNAs are vendor, researcher, open source, CERT, hosted service, bug bounty provider, and consortium organizations authorized by the CVE Program to assign CVE IDs to vulnerabilities and publish CVE Records within their own specific scopes of coverage." [24]

In simpler terms a CNA is an organization given coverage of a specific area or scope of products. The CNA is given the right to make and assign CVE IDs. Some examples of CNAs include organizations like: Microsoft Corporation, Google LLC, Netflix Inc, Meta Platforms, Apple Inc, Kernel.org, Fedora Project, and even Airbus.

When a security researcher wants to register a CVE they first need to find the appropriate CNA. In many cases this is an easy process. For example, if the vulnerability is found in a Google product, then the security researcher should send their form under Google's CNA. In other cases, finding the proper CNA might be difficult. In our own example, FydeOS is not a registered CNA, nor do they have any parent organization which is. In such cases where the security researcher is unable to determine the correct CNA, Mitre provides two Certified Numbering Authority of Last Resort (CNA-LR). These two CNAs serve as a catchall CNAs for cases in which a vulnerability doesn't fit under any known CNA's aegis. Each CNA-LR covers a different scope. "CISA ICS CNA-LR for industrial control systems and medical devices" and "MITRE CNA-LR for all vulnerabilities, and Open Source software product vulnerabilities, not already covered by a CNA listed on this website" [25] Since the vulnerability originates from OpenFyde, and open source product, and isn't covered by a known CNA, this vulnerability should be submitted to the "Mitre CNA-LR" CNA.

CWE - Common Weakness Enumeration

Mitre, not only identifies common vulnerability exposures, but also identifies common weakness types. These are called CWE or a common weakness enumeration. A CWE identifier is used to identify a particular common weakness. These identifiers are given the prefix "CWE" followed by an integer. For example, CWE-276 is the CWE ID that corresponds to Incorrect Default Permissions.

What makes something a weakness rather than a vulnerability? A vulnerability is created when one or more weaknesses in a product are in a context as to be exploited. A weakness is something that, under the right circumstances, could contribute or introduce a vulnerability. In other words the term vulnerability includes the context of a weakness and how it's exploited. A CWE is not, in itself, a vulnerability. CWEs act as a way to categorize vulnerabilities into types. This data can then be used to analyze trends in vulnerabilities. Sometimes a vulnerability might be made up of multiple CWEs working in conjunction.

Example:

In 2021, CVE-2021-30493 [26] and CVE-2021-30494 were found inside Razer Synapse, a companion program for Razer PC peripherals. When a Razer peripheral device, such as a Razer mouse, was plugged into a Windows 10/11 computer, Windows would automatically install the driver and launch the installer for the companion software, Razer Synapse. The software, during normal installation, prompts you to run it as an Administrator for you to install it; however, Windows automatically launches signed driver installations as Administrator. This privilege carried over to the Synapse installation who's installer was automatically launched during the process. From the Razer Synapse installation screen users were allowed to change Synapse's install location. This then opens Windows Explorer, which inherits the Administrator permissions. From Windows Explorer, users are able to right click in a folder and press "open powershell window here". This would launch a command prompt with Administrative rights.

The Razer Synapse vulnerability was linked to CWE-276: Incorrect Default Permissions. The vulnerability required, not only the CWE, but a number of other factors to coincide. One of these factors was that the Razer Synapse program allowed users to access Windows Explorer. While Razer Synapse shouldn't have been launched as Administrator, this CVE wouldn't exist if it hadn't allowed access to Windows Explorer. In this hypothetical, there would have been a CWE, but no CVE.

The taxonomy of vulnerabilities and weaknesses can be complex. To help organize related CWEs Mitre has them organized in a graph format. This allows CWEs to have parents, children, siblings, as well as assigned categories. For example, "CWE-1393: Use of Default Password" is a child of "CWE-1392: Use of Default Credentials". Selecting the underlying CWE for a given vulnerability is a process called Root Cause Mapping.[27] Root Cause Mapping can be a laborious process filled with many technicalities.

The vulnerability found in FydeOS falls into a few CWEs.

CWE-1392: Use of Default Credentials

CWE-1393: Use of Default Password

CWE-258: Empty Password in Configuration File

Now that we have both our CNA and our CWE identified, we have everything we need to fill out the CVE Request for ID forum. We'll soon receive an email telling us that our submission will be reviewed by our CNA, in this case a CVE Assignment Team member from Mitre. After the review process we'll receive another email detailing our submission and our brand new CVE identifier. This particular vulnerability was assigned the identifier CVE-2024-25825.

Mitre and the CVE program will keep this vulnerability undisclosed until a publication about it is released. This might be an announcement from the parent company, a post on twitter, or a thesis like this very paper.

**Retrospect**

It occurred to me, only after talks with the CISA, that the usual disclosure procedure might be actively harmful when dealing with threat like a nation state actor. Had FydeOS been connected to a nation state actor, approaching them with this vulnerability could have had consequences.

It would be prudent for all security professionals who are concerned by involvement by a nation state actor to consider going directly to their respective countries CERT before reaching out to the parent company. If the developers, organization, or parent company are compromised or a threat actor, going directly to them first would give them ample warning to prepare a defense against any investigation; making stopping future attacks more difficult.

**Incident Response and Mitigation**

FydeOS has told me that their next release will patch this vulnerability. Unfortunately auto-updates are only included for their paying customers (in their enterprise service plan), meaning that users will manually update by downloading and running a file. As such, only users who installed the new version, or manually updated will be patched.

It is possible to create a script capable of patching this vulnerability by using the vulnerability itself as an entry point. Unfortunately, the process of running a shell script is more complicated than updating your system, and only slightly less complicated than manually changing the password.

**Fyde Legal and Ethical Concerns**

Fyde's actions have some ethical concerns. They repeatedly assure their users of FydeOS's safety and security, while knowing there's a massive vulnerability, and provide no way to deploy security updates to their non paying user, who make up the vast majority of their user base.

The following is a quote form FydeOS's FAQ page:

> "Does FydeOS carry malware that might compromise my information security?
>
> We can only guarantee that operating system images distributed from official FydeOS download page bearing the FydeOS logo are secure, harmless and comply with our Privacy Policy and other legal provisions relating to Data Protection Act in the United Kingdom.
>
> We cannot guarantee the security of FydeOS-like operating system products distributed by third parties, derivative products based on FydeOS technology, or distributions based on open source projects maintained by Fyde Innovations.

> The Chromium OS open source project, on which FydeOS is based, takes security very seriously. If you become aware of a significant security risk in Chromium OS (and any underlying components that come with it) that has not been or has been disclosed, please do not hesitate to reach out for us."

[7]

Developers have a responsibility to be truthful to their users and to provide them with secure software. Fyde has failed at both of these. Aside from the ethical concerns, Fyde could have legal exposure. Their claim that "operating system images distributed from official FydeOS download page bearing the FydeOS logo are secure" is false.

## Investigation

I judged it prudent to conduct a more thorough investigation into FydeOS. Primarily, I wanted to compare and contrast the differences between the Chinese version and the Global version. My primary goal was to find these differences and search for any malware or added vulnerabilities that could be present in one, or both images of the operating system.

**Initial Enumeration**

Fyde has two primary sites, a .io domain, and a .com domain. Their .com domain services mainland China, while their .io domain services the rest of the world.

Note: In some of my documentation I label these as US and CH versions, this was for increased terminal usability rather than any geographical accuracy. More accurately, 'US' simply refers to the "non-Chinese" version. I found during my investigation that I was often confusing io and com so I had to change the naming schema to something different. I found CH and US to work well for me.

My first step was to determine if the .com (Chinese) version of the operating system was different from the .io (non-chinese) version. It stands to reason that if you're purposefully putting a vulnerability in an operating system, you might not want your own people exposed to that vulnerability.

**Delivery Differences**

The first difference to note is that they offer different download options. The .com site offers three download links. The first link is what appears to be a OneDrive link but doesn't come from a microsoft domain.

The first download link is hosted by sharepoint.cn. At first glance this link appears to be a genuine Microsoft OneDrive link, but after some investigation things get strange. Navigating to just the base url, without the specific link information, we find a very sparse landing page.

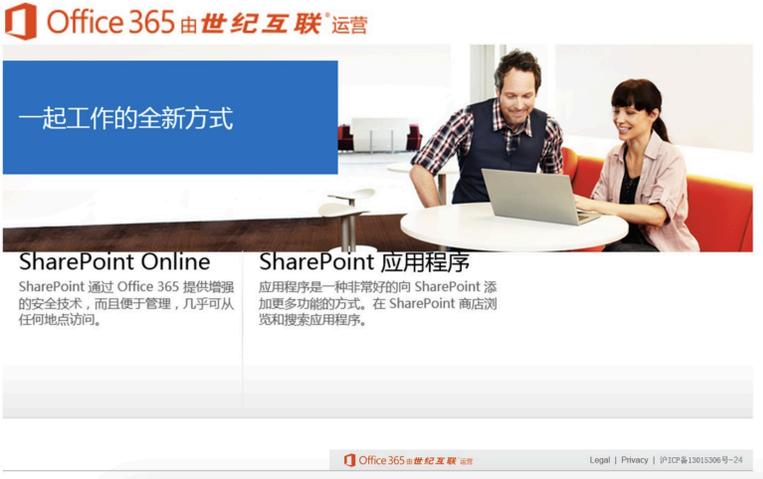

The landing page of sharepoint.cn

The entire page's source code is 35 lines and primarily consists of an image. In fact, they go out of their way to avoid using text. The footer that reads "Legal", "Privacy", and their Internet Content Provider license number, are all pictures, rather than text. It is, by far, one of the strangest ways of creating a website I've ever come across. The largest flaw however is the margin between the text "SharePoint Online" and the image above it, they're actually clipping into each other. The links in the footer, the ones that are pictures rather than text, are all dead links that couldn't resolve a host name.

Given this information, this appears to not be a genuine Microsoft website, despite styling itself like one. Fortunately a search on the domain "sharepoint.cn" explains some of the strangeness. It seems that Microsoft licenses their 365 office suite to a company called 21Vianet in China. From Microsoft's website: "The Microsoft 365 operated by 21Vianet version of Microsoft 365 is specific to China. Microsoft does not operate the service itself. 21Vianet operates, provides, and manages delivery of the service." Microsoft's website also informs us that "sharepoint.cn" is one of the domains registered to 21Vianet. [28]

The second link fydeos.com gives us directs us to an icloud.com link. This link seems rather straightforward and goes to a genuine icloud domain. The third link actually sends us to the fydeos.io (non-chinese) domain.

The .io links are much simpler but less exciting. The first link is directly from the fydeos.io website, with the second from icloud, and the third being a google drive link.

When placing all down download URLs into a table we find something interesting. The download links to the same download services, notably fydeos.io and icloud.com, offer different files despite being labeled as the same product. Most notably we notice the file name difference in the fydeos.io download links. These differences are highlighted below.

| Domain | fydeos.io (non-chinese) | fydeos.com (chinese) |
| --- | --- | --- |
| fydeos.io Downloads | https://download.fydeos.io/FydeOS_for_PC_v17.1-io-stable.img.xz | https://download.fydeos.io/FydeOS_for_PC_v17.1-com-stable.img.xz |
| icloud.com Downloads | https://www.icloud.com/iclouddrive/038BqUNt5WqzBxphg3DZNdN6Q#FydeOS_for_PC_v18.0-SP1-com-stable | https://www.icloud.com/iclouddrive/084xcKMp9Cm1l9Xy2wunW2NGw#FydeOS_for_PC_v18.0-SP1-io-stable |
| Other (Google Drive and Sharepoint) Downloads | https://drive.google.com/file/d/161XEBqMOfr8kltT8EIVBVZO0mdUB6AgM/view?usp=sharing | https://fydeos-my.sharepoint.cn/:u:/g/personal/fyde_fydeos_partner_onmschina_cn/EexN5UXim6JCppRMn-Wo6nwBJ |

| | | hKLfBhLhBuUuLA__rHPog?e=LmgeEG |
|---|---|---|

**Size Differences**

On both the .com and .io domains the operating system is delivered via a `.img.xz` file.

`FydeOS_for_PC_v17.1-io-stable.img.xz`

`FydeOS_for_PC_v17.1-com-stable.img.xz`

These files contain the "image" of the operating system. Let's break down what this .img.xz file extension means.

In computing a disk image is a single file that contains an exact copy of a disk or a partition's contents. Rather than storing just files, like as you would on a normal flash drive, disk image lets you capture and store things you'd normally not be able to, for example, a disk's file system. This makes it the perfect medium to store an operating system's installation media. An operating system's installation media usually require several partitions and are usually specialty designed operating systems themselves.

The ".img" extension is short for disk image, and implies that the file using that extension should be read as a disk image.

The .xz file extension implies a type of compression is used to reduce the file's size. Based on the Lempel–Ziv–Markov chain algorithm (LZMA), a type of lossless compression algorithm, the xz utilities software is a common choice for file compression.

The combination of .img.xz implies that it's an image file compressed with the LZMA algorithm. The .io (US) version is 1.6 GiB while the .com (CH) version is 2.0 GIB. Just by looking at the size we immediately know something is different between the two files, but they're still compressed. This discrepancy on its own is concerning.

When uncompressed both come out to be the same 6.8GiB. This is strange as both should be of similar size when compressed. As of present I still don't know what causes this discrepancy. Perhaps just differences in the LZMA compression algorithm.

**Image Hash Comparison**

Just like passwords, hashes can also be applied on files. Hashes are useful in determining if two files are identical as, even a small difference, will make a large difference in the resulting hash. In this case we're going to use the popular algorithm SHA256. First we uncompress the xz files into .img files, then we hash the files, and compare the hashes.

If the hash is the same, then it's safe to assume they're the same file down to the bit, but if the hashes are different then the files are different.

```
~/Downloads $ sha256sum CH-FydeOS_for_PC_apu_v17.1-com-stable.img
```

2d7126c4a109b055c61d5164d68454e1961f49a350811e57e1a41693a5ee0cdf

CH-FydeOS_for_PC_apu_v17.1-com-stable.img

```
~/Downloads $ sha256sum US-FydeOS_for_PC_apu_v17.1-io-stable.img
```

f4699f5229545a419a2a29538c40404e6592cc9b92f46279ea06f329327f06c4

US-FydeOS_for_PC_apu_v17.1-io-stable.img

As we can see, these two hashes (highlighted above), are different from one another. At this point we can confirm there's a difference between the US and CH versions. Now we need to figure out how they're different.

**User Experience Differences**

As an initial test I installed both the IO and COM versions on identical hardware. If there were any user experience differences, for example a different localization (language) as its default, this could explain the discrepancy we saw. Strangely enough, however, both were an identical install experience. Once installed, I found no visible differences between the experiences. Whatever the differences were between the two operating systems, it seemed to have nothing to do with user experience.

**Partition Difference**

      To narrow down my search, I wanted to determine which part of the installation images was different. As previously stated above, installation images usually consist of multiple partitions. In an over simplified explanation, partitions are a way of separating a disk into several virtual disks. This lets you install separate file systems on different partitions on the same disk.

The next step was to mount (attach) the .img files to my system, so I could observe their partitions.

The following is an edited list of commands used to enumerate these partitions along with some selected outputs.

```
$ losetup -P /dev/loop0
~/Downloads/US-FydeOS_for_PC_apu_v17.1-io-stable.img
```

```
$ losetup -P /dev/loop1
~/Downloads/CH-FydeOS_for_PC_apu_v17.1-io-stable.img
```

```
lsblk output + lsblk --fs
```

Here loop0 is a loopback device showing the IO version, while loop1 is mapped to the COM version.

```
NAME          MAJ:MIN RM   SIZE RO TYPE MOUNTPOINTS
loop0           7:0    0   6.8G  0 loop                         (US)
├─loop0p1     259:8    0     4G  0 part
├─loop0p2     259:9    0    16M  0 part
```

```
├─loop0p3    259:10   0    2.7G  0 part
├─loop0p4    259:11   0     16M  0 part
├─loop0p5    259:12   0      2M  0 part
├─loop0p6    259:13   0    512B  0 part
├─loop0p7    259:14   0    512B  0 part
├─loop0p8    259:15   0     16M  0 part
├─loop0p9    259:16   0    512B  0 part
├─loop0p10   259:17   0    512B  0 part
├─loop0p11   259:18   0      8M  0 part
└─loop0p12   259:19   0     32M  0 part
loop1          7:1    0    6.8G  0 loop                    (CH)
├─loop1p1    259:20   0      4G  0 part
├─loop1p2    259:21   0     16M  0 part
├─loop1p3    259:22   0    2.7G  0 part
├─loop1p4    259:23   0     16M  0 part
├─loop1p5    259:24   0      2M  0 part
├─loop1p6    259:25   0    512B  0 part
├─loop1p7    259:26   0    512B  0 part
├─loop1p8    259:27   0     16M  0 part
├─loop1p9    259:28   0    512B  0 part
├─loop1p10   259:29   0    512B  0 part
├─loop1p11   259:30   0      8M  0 part
└─loop1p12   259:31   0     32M  0 part
```

```
NAME          FSTYPE FSVER LABEL      UUID
FSAVAIL FSUSE% MOUNTPOINTS
loop0
(US)
├─loop0p1    ext4   1.0    STATE
c05442be-e437-4031-9dd1-46991eba02ab
├─loop0p2
├─loop0p3    ext2   1.0    ROOT-A
├─loop0p4
├─loop0p5
├─loop0p6
├─loop0p7
├─loop0p8    ext4   1.0    OEM
```

```
2495d5e4-c1ee-4956-8216-e3765fdc9446
├─loop0p9
├─loop0p10
├─loop0p11
└─loop0p12  vfat    FAT16 EFI-SYSTEM 29DE-F214
```

```
(CH)
├─loop1p1   ext4    1.0   STATE
bce8f284-a3d2-4080-91d7-aa28813e803e
├─loop1p2
├─loop1p3   ext2    1.0   ROOT-A
├─loop1p4
├─loop1p5
├─loop1p6
├─loop1p7
├─loop1p8   ext4    1.0   OEM
fcde8c30-72dd-4b21-b649-de55b0176009
├─loop1p9
├─loop1p10
├─loop1p11
└─loop1p12  vfat    FAT16 EFI-SYSTEM 5701-D840
```

We can see both are largely the same. They both have 12 partitions, 2 ext4 partitions, 1 ext2 partition, and a vfat partition all in the same order.

Since Fyde is based on Chromium, I've added what a normal ChromeOS Felx install image looks like. I'll be using this as a baseline.

Normal ChromeOS Flex Installer

```
NAME          MAJ:MIN RM   SIZE RO TYPE MOUNTPOINTS
sda           8:0      1    29G  0 disk
```

```
├─sda1         8:1    1      4G  0 part
├─sda2         8:2    1     16M  0 part
├─sda3         8:3    1    2.3G  0 part
├─sda4         8:4    1     16M  0 part
├─sda5         8:5    1      2M  0 part
├─sda6         8:6    1    512B  0 part
├─sda7         8:7    1    512B  0 part
├─sda8         8:8    1     16M  0 part
├─sda9         8:9    1    512B  0 part
├─sda10        8:10   1    512B  0 part
├─sda11        8:11   1    512B  0 part
└─sda12        8:12   1     64M  0 part
```

```
NAME         FSTYPE FSVER LABEL       UUID
FSAVAIL FSUSE% MOUNTPOINTS
sda
├─sda1       ext4   1.0   STATE
2b3f96a6-b134-4c2d-a476-8cf20dada4f5
├─sda2
├─sda3       ext4   1.0   ROOT-A
├─sda4
├─sda5
├─sda6
├─sda7
├─sda8       ext4   1.0   OEM
7ff34986-6e50-42d8-90c5-308535b4c4e6
├─sda9
├─sda10
├─sda11
└─sda12      vfat   FAT16 EFI-SYSTEM BE0E-84F6
```

We can note some differences between the standard ChromeOS Flex installer and the FydeOS installer, (the ext2 partition has switched to ext4) but as FydeOS is known to be a modified fork of ChromeOS, such drift is not unlikely.

Next was to determine if each partition was different. At this point, we're not sure if there's only a single change in a single partition or if every partition is different.

To find out which, we'll hash the contents of each partition individually and compare its hash with the corresponding pariton's hash of the other image.

```
~/Downloads   sudo dd if=/dev/loop0p1 bs=1M | sha256sum
        ✔
4096+1 records in
4096+1 records out
4295023104 bytes (4.3 GB, 4.0 GiB) copied, 3.24077 s, 1.3 GB/s
71adec86dd1538d0ddc4ddb52a124d2568802024a939a9b5c46d6f790b4d61ac  -
    ~/Downloads   sudo dd if=/dev/loop1p1 bs=1M | sha256sum
        ✔   3s
4096+1 records in
4096+1 records out
4295023104 bytes (4.3 GB, 4.0 GiB) copied, 3.49673 s, 1.2 GB/s
e411438bf510da5261b45b225668ad2726c0b535f3873a7319cf358c7754b9f3  -
```

As we can see from the output above, the first partition is different on both images.

```
~/Downloads   sudo dd if=/dev/loop0p2 bs=1M | sha256sum
        ✔   4s
16+0 records in
16+0 records out
16777216 bytes (17 MB, 16 MiB) copied, 0.024983 s, 672 MB/s
f1f06591ccab09d58b590fb4a8f403aa9b941c419c7103610510ac8bf0db6af5  -
    ~/Downloads   sudo dd if=/dev/loop1p2 bs=1M | sha256sum
        ✔
16+0 records in
16+0 records out
```

```
16777216 bytes (17 MB, 16 MiB) copied, 0.0158329 s, 1.1 GB/s
01964c0bb09c97d14c6545da0da6acf020a87e1c6624b85159687d51918b463b  -
```

```
~/Downloads  sudo dd if=/dev/loop0p3 bs=1M | sha256sum
        ✔
2724+0 records in
2724+0 records out
2856321024 bytes (2.9 GB, 2.7 GiB) copied, 2.66567 s, 1.1 GB/s
62e87dda7ea6c276a49286fc4a6e28ca15bec56248dbef3bffbfd6913518d3de  -
    ~/Downloads  sudo dd if=/dev/loop1p3 bs=1M | sha256sum
        ✔
2724+0 records in
2724+0 records out
2856321024 bytes (2.9 GB, 2.7 GiB) copied, 2.33471 s, 1.2 GB/s
45e6b0ab56ff2d5edd6d1cf948fdf2e289327e97cdf393c26b570e321d82296d  -
```

This pattern of mismatching hashes continues for all 12 partitions. At this point we can be assured that whatever difference there is between both the IO and COM images, persists across all partitions.

**Binary Analysis Differences**

Next is narrowing down these differences. To help with this I've created a script. The pseudo code is as follow:

```
function compare(filesystem A, filesystem B):
# First we gather the names of all the files on a filesystem
List filesA = List all the files on file system A
```

```
List filesB = List all the files on file system B
# next we determine if there are any differences in the number and
name of files between the two files systems
List diff = diff(filesA, filesB)
# here we hash the files found
List hashA, hashB
For file in filesA:
    filesA.append(hash(file))
For file in filesB:
    filesB.append(hash(file))
#lastly we compare the hashes between the hash lists
List diff(hashA hashB)
# by using this list we're able to search up which files are
different (note: in the actual script I use tuples to keep track of
which hash corresponds to which file). This example shows a
simplified version
```

The script's resulting output reported that most of the files were the same, with the exclusion of some binaries, and most worrying, the kernel. The kernel is one of the most important parts of any operating system. Often called the "heart" of an operating system, it's responsible for interfacing with a system's hardware, scheduling how much compute time programs get, and running things like drivers. Malware that resides in the kernel, called a kernel-mode rootkit, is extremely difficult to detect and remove.

"Kernel mode rootkits are extremely dangerous to the runtime integrity of a Linux system. These rootkits have the power to add, delete, or modify any data that kernel or userland applications request about the state of the system. This can include information such as lists of running processes, loaded kernel modules, active network connections, files within a directory, and even the contents of those files." ~ The Art of Memory Forensics [29]

Kernel-mode rootkits constitute a larger threat than loose root access.

**Kernel Binary Analysis**

Both the IO and COM versions of FydeOS has two kernels, labeled vmlinuz.A and vmlinuz.B. These are compressed kernel images which the bootloader will move into RAM during startup of the system. This method of having two, redundant, kernels is a common method to insure that if one is somehow corrupted (e.g. like during power loss during an update), the system has a fallback kernel to boot to. Both the IO and COM versions defaulted to using vmlinuz.A and so I focused my investigation on that kernel.

My first step was to take the binary files that are the kernels and convert them into a more readable format, in this case, hexadecimal.

Hexadecimal, often called hex, is a numbering system that can accurately represent binary with no remainder, so it's often used to analyze binary files. Where the binary number system you're familiar with starts at zero and ends at nine, before adding a new place, hex starts at zero, goes past nine, and includes the letters A, B, C, D, E, F after. In the normal decimal system we have ten total digits, 0-9. In hex we have 16 total digits, 0-F. Binary is made up of ones and zero, these are called bits. A byte in binary is eight bits. Four bits can represent sixteen numbers (0000 = 0 while 1111 = 16). By using hex, we can communicate the state of a byte by using only two "digits" between the numbers 0-F.

| Binary | Decimal | Hex |
| --- | --- | --- |
| 0000 0000 | 0 | 00 |
| 1111 1111 | 255 | FF |
| 0001 0001 | 17 | 11 |
| 1001 1011 | 155 | 9B |

While decimals fluctuate between one and three places to represent a binary value of eight places, a hex will always stay just two places, since its carryover value aligns with the max value of half of a byte (called a nibble).

After converting to hex, we then compare the two files.

```
$   ~/Downloads   hexdump US/EFI/syslinux/vmlinuz.A -C > USKernel.hex
$   ~/Downloads   hexdump CH/EFI/syslinux/vmlinuz.A -C > CHKernel.hex

$   ~/Downloads   diff USKernel.hex CHKernel.hex > Kernel.diff
```

And… we get a ton of, what at first glance, looks like gibberish.

Note: The font is reduced to increase readability and to keep the original formatting.

```
< 000000a0  19 02 00 00 00 00 00 00  00 00 00 7f 89 00 00 02  |................|
---
> 000000a0  19 02 00 00 00 00 00 00  00 00 10 7f 89 00 00 02  |................|
37c37
< 00000240  00 00 00 00 00 00 00 00  a8 03 00 00 14 c5 88 00  |................|
---
```

```
> 00000240  00 00 00 00 00 00 00 00  a8 03 00 00 1a c5 88 00  |................|
39c39
< 00000260  00 c0 f9 01 90 01 00 00  9c 94 89 00 8c d8 8e c0  |................|
---
> 00000260  00 c0 f9 01 90 01 00 00  ac 94 89 00 8c d8 8e c0  |................|
983c983
< 00003d80  20 31 35 20 30 37 3a 32  35 3a 33 36 20 55 54 43  | 15 07:25:36 UTC|
---
> 00003d80  20 31 35 20 30 37 3a 33  37 3a 33 38 20 55 54 43  | 15 07:37:38 UTC|
1005c1005
< 00004010  0d 8d 85 80 a7 89 00 89  40 02 0f 01 10 b8 18 00  |........@.......|
---
```

This continues for 6,931,740 lines. What's interesting is that some of the hexadecimals that are appearing are mapped to ASCII characters. ASCII, or the American Standard Code for Information Interchange, is a standard way of encoding letters as binary. You can see the ASCII bracketed by `|` on the right hand side of the output.

Of particular importance are these two lines which appear to resolve into timestamps.

```
< 00003d80  20 31 35 20 30 37 3a 32  35 3a 33 36 20 55 54 43  | 15 07:25:36 UTC|
---
> 00003d80  20 31 35 20 30 37 3a 33  37 3a 33 38 20 55 54 43  | 15 07:37:38 UTC|
```

At this point I felt it prudent to see what other text might be encoded into the kernel. To do this we use a command called `strings`, which will only return characters who's bytes are valid and printable ASCII. Many bytes however, can be valid ASCII without being intended to be read as such. A quick scroll through the data revealed some interesting metadata encoded into the kernel. I've filtered out the gibberish and isolated the metadata by using the `grep` command, the results of which are below. For reference I've also included the metadata found in a normal ChromeOS Flex kernel.

Global - IO Output

```
~/Downloads $   strings US/EFI/syslinux/vmlinuz.A| grep built
5.15.108-18907-gba143be42d3a-dirty (builty@fydebeast) #2 SMP PREEMPT
Wed Nov 15 07:25:36 UTC 2023
```

China - COM Output

```
~/Downloads $   strings CH/EFI/syslinux/vmlinuz.A| grep built
5.15.108-18907-gba143be42d3a-dirty (builty@fydebeast) #2 SMP PREEMPT
Wed Nov 15 07:37:38 UTC 2023
```

ChromeOS Flex

```
~/Downloads/Flex/syslinux $   strings vmlinuz.A | grep kernel
5.15.140-21046-g921d2194f426 (cros-kernel@chromium.org) #1 SMP
PREEMPT Wed, 7 Feb 2024 21:32:19 +0000
```

First we see the version, in this case both IO and COM are on versions 5.15.108. FydeOS is several revisions behind the ChromeOS version, but considering their build frequency is less than that of FydeOS, this is to be expected.

Next we see details of who built the kernel. In this case, this kernel was compiled by the user `builty` on a machine called `fydebeast`. Lastly we see the timestamp differences. As we can see the COM (Chinese) version was built exactly 12 minutes and 2 seconds after the IO (Global) version.

While some, if not most, of the difference between the two kernels is related to metadata, I can't confirm if that's all the difference there is. There is, however, a quick and dirty enumeration solution. If a kernel-mode rootkit were introduced we'd expect the kernel to be larger in size, or at least a different size. While there are methods of introducing malware without changing the kernel's size, it would require modification at the binary level. Such malware is extremely difficult to produce and I find it unlikely that it'd be used on an operating system such as this.

To find the size of the kernels well use a tool called `du`, short for Disk Usage. This tool, when given the right arguments, can tell use the size of the kernel down to the byte.

Kernel Size in Bytes

Global - IO Output

```
$   ~/Downloads   du -b US/EFI/syslinux/vmlinuz.A
9038400 US/EFI/syslinux/vmlinuz.A
```

```
$   ~/Downloads   du -b US/EFI/syslinux/vmlinuz.A
```

```
9038400 US/EFI/syslinux/vmlinuz.A
```

China - COM Output

```
$   ~/Downloads   du -b CH/EFI/syslinux/vmlinuz.B
9038400 CH/EFI/syslinux/vmlinuz.B
```

```
$   ~/Downloads   du -b CH/EFI/syslinux/vmlinuz.A
9038400 CH/EFI/syslinux/vmlinuz.A
```

As we can see, to the byte, these kernels are the same size. While we can't be one hundred percent sure without pouring over the entire kernel's binary code, I believe it safe to assume that there's no kernel-mode rootkit in the global version that's not present in the Chinese version.

Lastly, I repeated this process for the kernel's modules. Kernel modules are basically option kernels addons which load dynamically, as needed, during startup. They're what allow you to download a new driver without downloading a whole new kernel.

**Investigation Final Steps**

To finish my investigation I manually checked some of the more sensitive configuration files in the /etc/ directory, along with some of the binaries in the /sbin/ and /bin directories. These directories are responsible for containing executable programs which are often called by the operating system. Lastly I compared all the binaries with SUID, SGID, and GUID sticky bits and found them to have the same permissions. The SUID, SGID, and GUID are special permissions called sticky bits. They can modify the behavior of an executable or directory. For example, the SUID sticky bit is used to change the behavior of an executable such that it runs with the owner's permission. This can lead to vulnerabilities if it's not implemented properly.

**Fyde's Explanation**

Now that I've independently reviewed FydeOS, I decided to reach out to my previous contact at Fyde, @Alpha (fydeos_alpha). The choice to contact them after my investigation was done to avoid confirmation bias.

> Me: "If you don't mind me asking, what's the difference between the US and Chinese versions? More specifically the com vs io like in FydeOS_for_PC_apu_v17.1-com-stable.img vs FydeOS_for_PC_apu_v17.1-io-stable.img"

> fydeos_alpha: "Oh hey, hope all is well. Technically it's the Chinese version and the non-Chinese version. We use .com and .io to distinguish. The non-Chinese version has all the internet services deployed in the UK, it's also run by a UK company entity, so it's under the jurisdiction of the United Kingdom (and EU) in terms of data protection laws and regs.
>
> Whilst the .com version has everything within China, the laws and regs are just different. There are also some features only available in the .com version that are primarily designed to work in the connectivity within China

> here, https://fydeos.io/datageo/, has all the details
>
> Apparently the combination of GDPR+UK Data Protection Act is a nightmare for large companies, the rules are a lot more complicated than in the US, and anywhere else in the world, so i was told
>
> BTW, in a few days we will release FydeOS v18, where we have changed how the images are shipped. It will be in airtight mode (base image) by default where users will NOT get access to shell. There will be a button to press and some warning messages to read before users can switch the OS into developer mode so shell access can be restored
>
> sshd and port 22 will also be disabled by default"

At the time of writing, I can't find any evidence disputing fydeos_alpha's claim. The explanation provided by fydeos_alpha could explain the discrepancy between the systems.

**Conclusion**

FydeOS, a fork of ChromiumOS, is a web-centric distribution that promises security, privacy, and ease of use. While investigating the operating system to see if it'd be suitable for personal use, I came across a vulnerability. I followed procedure and disclosed this vulnerability to the relevant authorities. During this process, concerns of it being an intentional move by a nation state actor were brought up. Further investigation showed that it's unlikely to be a state sponsored vulnerability.

**Lessons Learned**

The typical process of responsible disclosure might have exceptions, depending on context. During such times in which the parent company is a potential suspect, it would behoove security researchers to skip reaching out to the parent company, and report their concerns directly to their countries computer emergency response team. Responsible disclosure isn't the only form of disclosure, and there may be legitimate ethical reasons for modifying responsible disclosure.

There's an assumption by many computer enthusiasts that, just because software is open source, means that it's regularly audited for vulnerabilities. The story of FydeOS's first CVE is a good example on why this isn't true. While software being open sources does mean that people can freely audit that software, it doesn't mean people will audit that software. This vulnerability should have been found a lot sooner. It should never have gotten to this point. Open source software, doesn't mean it's safe from having obvious vulnerabilities, especially when there's no one going through the trouble to audit the code.

FydeOS's response to this vulnerability left much to be desired. From their response I believe they don't have a comprehensive security policy. As a large open source project, user safety should be high among their concerns. I believe it would help FydeOS to invest time in developing a security policy and strategy, along with an incident response plan, as many other distributions of their size have.

Lastly, it's important to know that you can make a difference in the area of cyber security. This CVE wasn't challenging to find. Many amateur computer enthusiasts could have found and identified it. I'm sure I'm not the first person to notice this vulnerability, only the first to report it. If you find a vulnerability, report it. You can make a difference.

**Future Work**

There are areas in which this investigation can be continued. Notably, analyzing the network traffic from the IO and COM versions and comparing them to fydeos_alpha's claims. At this time we still don't know the cause of the differing compressed file sizes. This could warrant further investigation. FydeOS seems to have departed from a few of ChromeOS's default security measures. It would be worth investigating if any other security measures are missing, such as the use of TMP and disk encryption.

# Appendix A - Quotes from FydeOS

## A.1 Security Assurance in FydeOS

"FydeOS is engineered with security as a priority, making it a reliable choice for your daily computing needs. Unlike traditional Windows PCs, FydeOS eliminates the necessity for anti-virus software or frequent malware scans, offering a seamless and worry-free experience." [6]

## A.2 Official Image Distribution Guarantee

"We can only guarantee that operating system images distributed from official FydeOS download page bearing the FydeOS logo are secure, harmless and comply with our Privacy Policy and other legal provisions relating to Data Protection Act in the United Kingdom." [7]

## A.3 Personal Information Protection Principles

"We are committed to maintaining your trust in us by adhering to the following principles to protect your personal information:

the principle of consistent authority and responsibility

the principle of clear purpose

the principle of choice of consent

the principle of minimum necessary

the principle of ensuring the security

the principle of subject participation

At the same time, we are committed to protecting your personal information with appropriate security measures in accordance with proven industry security standards." [30]

**Appendix B - Definitions**

Operating System (OS): An operating system is software that manages a computer's resources, provides various services to applications, and often provides a graphical user interface to interact (launch, exit, manage) with programs and files. It acts as an intermediary between the user, applications, and the underlying hardware components of a computer system. The primary functions of an OS include process management, memory management, file management, device management, IO management, security, and resource allocation. Popular operating systems include Microsoft Windows, macOS, Linux, IOS, and Android.

Linux: Linux is an open-source, Unix-like, POSIX compliant, operating system based on the Linux kernel, which was initially developed by Linus Torvalds in 1991. It is a free and customizable OS that runs on various hardware platforms. Linux's source code is available under the GNU General Public License (GPL), allowing developers to modify and distribute their own versions of the operating system.

Responsible Disclosure: Responsible disclosure, also known as coordinated disclosure or informed disclosure, is the process of reporting vulnerabilities in software, hardware, or network infrastructure to the relevant parties (e.g., vendors, developers) while adhering to ethical and legal guidelines. This practice involves responsible security researchers who privately notify affected organizations about identified vulnerabilities before making their findings public. The main goal of responsible disclosure is to give the organization an opportunity to address the issue without exposing it to potential exploitation by malicious actors, thus minimizing risks to users and systems. Responsible disclosure typically follows a predefined timeline for patch development and implementation, with periodic updates provided to all stakeholders involved in the process.

Common Vulnerability Exposure (CVE): A CVE is an identifier assigned by the Mitre Corporation's Common Vulnerabilities and Exposures list to an instance or report of a vulnerability in software, hardware, or network infrastructure. The purpose of CVEs is to provide a standardized nomenclature for referencing specific security issues, enabling organizations to track, analyze, and manage vulnerabilities more efficiently. Each CVE entry contains information such as the affected product, version, description, impact, and references to related advisories or patches. By using CVE identifiers, cybersecurity professionals can streamline communication about potential threats and facilitate collaboration between vendors, researchers, and security practitioners in addressing and mitigating vulnerabilities across various systems and platforms.

ChromeOS: A lightweight, Gentoo (Linux) based operating system, developed by Google. Built around the Chrome web browser, its primary purpose is to deliver a user-friendly experience in its delivery of web applications. With a strong emphasis on security, ChromeOS is considered "secure by default" and has features such as automatic updates, verified boot (TPM authentication), sandboxing, built-in malware detection, built in encryption, and limited user privileges. This makes it an attractive offering for many enterprise use cases. It is, however, bound to the Google ecosystem, which is known for its data collection. It has high security, but low privacy. While largely based on ChromiumOS, it's open source cousin, ChromeOS contains some closed source code and proprietary software. d

ChromiumOS: A open source (BSD and GPL) operating system run by The Chromium Project under Google. ChromiumOS often acts as an upstream for ChromeOS's open source parts. While both share a similar foundation, ChromiumOS represents the pure, community-driven project while ChromeOS adds proprietary elements and specific optimizations for certain devices and services provided by Google. ChromiumOS is also considered "secure by default" and ChromeOS inherits most of it's security features from ChromiumOS.

FydeOS: FydeOS is a custom distribution of ChromiumOS. It's designed to be a drop-in replacement for ChromeOS. The main difference between FydeOS and standard ChromiumOS lies in its departure from the Google ecosystem, while seeking feature parity with ChromeOS. It's "de-googlefication" makes it appealing to those who want to get away from Google's data collection.

OpenFyde: OpenFyde is to FydeOS what ChromiumOS is to ChromeOS. It's an open source, community driven upstream of OpenFyde. It forgos many of FydeOS's proprietary features, but is open source.

# Appendix C - Conversation with "fydeos_alpha"

Note: At this time I believed that there was another vulnerability with FydeOS's SSH configuration. I was mistaken. The SSH configuration allowed for root sign in, but didn't allow for sign in with passwords.

======= START OF TRANSCRIPT =======

fydeos-alpha — 05/02/2024 06:26

hey

Umbrarum — 05/02/2024 06:27

Hey. I have some quick questions before I get into what I found and why it goes beyond what you might assume. (edited)

fydeos-alpha — 05/02/2024 06:27

fire away please

Umbrarum — 05/02/2024 06:28

Are you an employee, a contractor, or a volunteer for Fyde Innovations?

fydeos-alpha — 05/02/2024 06:29

I am the co-founder of the company that makes FydeOS and openFyde, I am also the CEO

[06:29]

My name is Alpha, nice to meet you

Umbrarum — 05/02/2024 06:44

Nice to meet you too. I'm very concerned about the security of the OS, as I'll get into in a second. Both OpenFyde and FydeOS for PC are posed as alternatives to ChromeOS. The issue is that ChromeOS is rather secure by default, Fyde is not. You have a large user base; it's frankly inappropriate to no explicitly inform the user that OpenFyde and FydeOS for PC come in dev mode and that they're "not air tight and welcomes hacking". You have users who trust that their information is secure while using your OS. You have users who sign into their Google accounts, and bank accounts with your OS. You have, in fact, introduced many vulnerabilities where there were none before. Your configurations are so bad that I legitimately think they were introduced to purposefully prey on your users.

I don't think I can stress enough how big of an issue this is. I'll be blunt. It's a national security risk for people to use your software. There's a two part problem, the first is the vulnerabilities, the second is the fact that your user base isn't informed. While government employees are supposed to use gov systems, many use their own for their personal life, and opsec is never a sure thing. You might have users that hold confidential or classified IP on your systems.

I'll be blunt. I feel obligated to write a report to US-CERT/CISA, to the FBI, the EFF, and several other cyber security organisations.

[06:45]

This is a gross breach of trust of your user base.

fydeos-alpha — 05/02/2024 06:46

First, thanks for the write up mate. When you mentioned "introduced many vulnerabilities", are there any specific examples that you care to share?

[06:48]

And also "they were introduced to purposefully prey on your users", this is quite a strong allegation. Whilst I won't defend ourselves as yet, but I'd be keen to learn from you what could we do better to address your concerns

Umbrarum — 05/02/2024 06:53

Your /etc/shadow file for starters. It has a wildcard for the root password. In other words, anyone can just log into the root user with no password, even when not logged on due to ctl + alt + f2. It means anyone with physical acess to the system can just bypass all security measures.

Second is that your /etc/ssh/sshd.conf file allows root sign in by default, which mean, not only do you have no password for root, but you allow anyone to remotely sign into the system with no password.

I've not been able to fully test this but from what I've seen ssh is an enabled service by default.

This means, anyone, at anytime, can access all the data on the OS. Files, web history, anything. All they need is your IP. (edited)

[06:55]

It's difficult to accidentally make your configurations this vulnerable. I'm not sure you could call it a true vulnerability, its more like an open door. It's so bad that legally, anyone who accesses this information would have the legal right to do so, because it's competently insecure.

fydeos-alpha — 05/02/2024 07:02

It's true that we have left the shell access open for all public releases of FydeOS, including FydeOS for PC, for You and all openFyde builds. This is the default behaviour for dev images that are configured by the Chromium OS SDK. We could talk more on this later.

However, sshd on FydeOS will reject all incoming connections BEFORE users have explicitly configured a root password. So I'd defend that "anyone, at anytime, can access all the data on the OS. Files, web history, anything. All they need is the IP." is simply not true.

Admittedly, publicly available FydeOS releases aren't as secure as ChromeOS on Chromebooks on a default setting, but it should behave the same when a chromebook turned into dev mode. I agree that we have not publicly emphasis the difference between FydeOS and default ChromeOS in terms security, this is something we will definitely look into and improve.

Umbrarum — 05/02/2024 07:14

I apologise. I was panicing, upset, and angry at everything I found and I skipped over some of the configuration file. You are correct in that PasswordAuthentication is set to no. You have my apologies for that. But there are definitly A LOT of risks that need to be addressed.

[07:15]

And that your users should be aware of.

[07:15]

I know of NO linux distributions that ships with a wildcard root password. I'm sure they're out there, but they're not as heavily used as yours. (edited)

fydeos-alpha — 05/02/2024 07:18

Coincidentally we have planned to add a new release channel in the coming release, under the FydeOS Enterprise Solution, which will be made publicly available too, something called "FydeOS in verified mode". Images in the channel will be built as "base image", it will not allow users to 1)gain shell access 2)sideload apk 3)change to dev mode. It'd be the same behaviour as chromebook by default. This is how we approach most of our enterprise users, we thought it may be useful for some users too, especially parents, as we are ready to add guardian mode back in.

Publically available FydeOS images require users to manually format hard drives, flash the bootable images to the hardware to function. We believe it'd be secure enough for daily usage, given that most malware and virus aren't functional. Before users sign-in to their account using password, user files are encrypted and won't be accessible even if the intruder has root access to the rootfs.

I'd like to thank you for pointing things out to us, to me. I will 100% look into your suggestions and see what we can do better in this

[07:20]

On a side note, what are your suggestions on how we should deal with the root password situation in FydeOS? What will you do?

Umbrarum — 05/02/2024 07:30

I apologise again for my overreaction. I was very upset (and sleep deprived), but seeing how professionally you handled this I feel ashamed at how I handled myself. Again, please accept my apologies. The easiest solution would be a post install script that runs before shutdown of the initial installer. It'd edit the /etc/shadow file to remove the root password, then make some changes to the SSH config (along with any other security config changes). You should also have the SSH service disabled by default. This is the easiest solution but not the 'correct solution'; which would be editing the configs from their source before the instillation. I'm not super familiar with your build process but that might be as simple as an addendum to the overlays or creating a fork of ChromiumOS or at least their configs and adding them to the build process.

[07:34]

Ideally there should also be a way to streamline security updates. Some way to prompt the user that an important update is available, or allow them to opt-into auto security updates.

Umbrarum — 05/02/2024 07:38

For your new release channel, when 'changing to dev mode', the user should have sudo, but the root password should remain blank. If they need root access they can run sudo passwd root. SSH should also remain disabled by default with an option to enable.

Umbrarum — 05/02/2024 08:18

I love the project but I just don't have the time to work on this unpayed. I'm going to ask some of my associates if they'd like to look over it. I think some of them might be interested though. Any feedback I get from them I'll send to you.

fydeos-alpha — 05/02/2024 08:46

Hey sorry was held up by something, but wow thanks for the write up. I will definitely get someone to review these with me as we are still in the final planning stage of the coming major release. Once again, thank you for bringing these to me and the conversation. x

7 February 2024

Umbrarum — 07/02/2024 10:24

Do you have any plans to go public with this vulnerability before the next release? (edited)

fydeos-alpha — 07/02/2024 23:48

We are going to publish a FAQ article titled "Is FydeOS secure?" to detail the shell access and dev mode situation.

1 March 2024

Umbrarum — 01/03/2024 03:46

If you don't mind me asking, what's the difference between the US and Chinese versions?

[03:47]

More specifically the com vs io like in FydeOS_for_PC_apu_v17.1-com-stable.img vs FydeOS_for_PC_apu_v17.1-io-stable.img

fydeos-alpha — 01/03/2024 03:57

Oh hey, hope all is well.

Technically it's the Chinese version and the non-Chinese version. We use .com and .io to distinguish. The non-Chinese version has all the internet services deployed in the UK, it's also run by a UK company entity, so it's under the jurisdiction of the United Kingdom (and EU) in terms of data protection laws and regs.

Whilst the .com version has everything within China, the laws and regs are just different. There are also some features only available in the .com version that are primarily designed to work in the connectivity within China

[03:57]

here, https://fydeos.io/datageo/, has all the details

FydeOS

Alpha

Data Geoloacation preference - FydeOS

Data geolocation option for FydeOS services Last edited: 19 Apr 2023

What are data geolocations?

As part of our ongoing commitment to delivering better service and increased data privacy options for our expanding global user base, we will establish a new data centre in the United Kingdom. This facility

Umbrarum — 01/03/2024 03:58

Ahh apologies! I've been labelling them US and CH for ease of use and it just caught on in my head.

[03:59]

Thank you for your quick feedback btw!

fydeos-alpha — 01/03/2024 03:59

Apparently the combination of GDPR+UK Data Protection Act is a nightmare for large companies, the rules are a lot more complicated than in the US, and anywhere else in the world, so i was told

[04:02]

BTW, in a few days we will release FydeOS v18, where we have changed how the images are shipped. It will be in airtight mode (base image) by default where users will NOT get access to shell. There will be a button to press and some warning messages to read before users can switch the OS into developer mode so shell access can be restored

[04:02]

sshd and port 22 will also be disabled by default (edited)

Umbrarum — 01/03/2024 04:07

Awesome! I'm very glad to hear it!

======= END OF TRANSCRIPT =======

https://learn.microsoft.com/en-us/office365/servicedescriptions/office-365-platform-service-description/microsoft-365-operated-by-21vianet

[29]   M. H. Ligh, A. Case, J. Levy, and A. Walters, *The art of memory forensics: detecting malware and threats in Windows, Linux, and Mac memory*. Indianapolis, IN: Wiley, 2014.

[30]   Alpha, "Privacy," FydeOS. Accessed: Mar. 15, 2024. [Online]. Available: https://fydeos.io/privacy/
97